\documentclass[pdflatex,sn-mathphys-num]{sn-jnl}

\usepackage{graphicx}%
\usepackage{multirow}%
\usepackage{amsmath,amssymb,amsfonts}%
\usepackage{amsthm}%
\usepackage{mathrsfs}%
\usepackage[title]{appendix}%
\usepackage{xcolor}%
\usepackage{textcomp}%
\usepackage{manyfoot}%
\usepackage{booktabs}%
\usepackage{algorithm}%
\usepackage{algorithmicx}%
\usepackage{algpseudocode}%
\usepackage{listings}%

\begin{document}

\title[Article Title]{Pomeranchuk Instability Induced by an Emergent Higher-Order van Hove Singularity on the Distorted Kagome Surface of Co$_3$Sn$_2$S$_2$}

\author[1]{\fnm{Pranab} \sur{Kumar Nag}}
\equalcont{These authors contributed equally to this work.}
\author[1]{\fnm{Rajib} \sur{Batabyal}}
\equalcont{These authors contributed equally to this work.}
\author[2]{\fnm{Julian} \sur{Ingham}}
\equalcont{These authors contributed equally to this work.}
\author[1]{\fnm{Noam} \sur{Morali}}
\equalcont{These authors contributed equally to this work.}
\author[1]{\fnm{Hengxin} \sur{Tan}}
\author[1]{\fnm{Jahyun} \sur{Koo}}
\author[3]{\fnm{Armando} \sur{Consiglio}}
\author[4]{\fnm{Enke} \sur{Liu}}
\author[1]{\fnm{Nurit} \sur{Avraham}}
\author[2]{\fnm{Raquel} \sur{Queiroz}}
\author[3]{\fnm{Ronny} \sur{Thomale}}
\author[1]{\fnm{Binghai} \sur{Yan}}
\author[4]{\fnm{Claudia} \sur{Felser}}
\author*[1]{\fnm{Haim} \sur{Beidenkopf}}\email{haim.beidenkopf@weizmann.ac.il}

\affil*[1]{\orgdiv{Condensed Matter Physics}, \orgname{Weizmann Institute of Science}, \orgaddress{\city{Rehovot}, \postcode{7610001}, \country{Israel}}}

\affil*[2]{\orgdiv{Department of Physics}, \orgname{Columbia University}, \orgaddress{\city{New York}, \postcode{10027}, \country{USA}}}

\affil[3]{\orgdiv{Institute for Theoretical Physics and Astrophysics}, \orgname{University of Würzburg}, \orgaddress{ \city{Würzburg}, \postcode{D-97074}, \country{Germany}}}
\affil[4]{ \orgname{Max Planck Institute for Chemical Physics of Solids}, \orgaddress{\city{Dresden}, \postcode{D-01187}, \country{Germany}}}


\abstract{Materials hosting flat bands at the vicinity of the Fermi level promote exotic symmetry broken states. Common to many of these are van Hove singularities at saddle points of the dispersion or even higher-order van Hove singularities where the dispersion is flattened further. The band structure of kagome metals hosts both a flat band and two regular saddle points flanking a Dirac node. We investigate the kagome ferromagnetic metal Co$_3$Sn$_2$S$_2$ using scanning tunneling spectroscopy. We identify a new mechanism by which a triangular distortion on its kagome Co$_3$Sn surface termination considerably flattens the saddle point dispersion, and induces an isolated higher-order van Hove singularity (HOvHS) with algebraically divergent density of states pinned to the Fermi energy. The distortion-induced HOvHS precipitates a Pomeranchuk instability of the Fermi surface, resulting in the formation of a series of nematic electronic states. We visualize the nematic order across an energy shell of about 100 meV in both real-, reciprocal-, and momentum-spaces, as a cascade of wavefunction distributions which spontaneously break the remaining rotational symmetry of the underlying distorted kagome lattice, without generating any additional translational symmetry breaking. It signifies the spontaneous removal of a subset of saddle points from the Fermi energy to lower energies. By tracking the electronic wavefunction structure across the deformed Fermi surface we further identify a charge pumping-like evolution of the wavefunction center of mass. The mechanism we find for the generation of higher-order saddle points under a kagome distortion may be common to other kagome materials, and potentially other lattice structures, suggesting a generic new avenue for inducing unconventional electronic instabilities towards exotic states of matter.}


\maketitle

\section{Introduction}\label{sec1}

Recent years have seen increasing ability to manipulate band structures and induce flat bands in the vicinity of the Fermi energy. Ubiquitous to many of these systems are saddle points in the dispersion which generate van Hove singularities (vHS) in the density of states~\cite{PhysRev.89.1189}. Near these points, the density of states is logarithmically divergent, precipitating the appearance of exotic electronic states of matter with spontaneously broken symmetries. There are a few known mechanisms that further squeeze the dispersion around such saddle points via inducing adjacent gaps in the band structure. This enhances the logarithmic divergence into a higher-order vHS (HOvHS) with an algebraically divergent density of states, at which the leading momentum dependence of the dispersion is of quartic or even higher power \cite{Shtyk2017, NatComYuan, Chandrasekaran2020, Classen2020, Chandrasekaran2023a, Chandrasekaran2023b}. The HOvHS breaks the weak coupling dominance of superconducting pairing over density wave and nematic orders, thus promoting high-temperature realizations of exotic electronic order~\cite{Classen2020,Han2023}. A prime example is given by magic angle twisted bilayer graphene, in which hybridization of the Dirac bands from the two layers produces minibands with a significantly reduced bandwidth. As the hybridization gaps increase they progressively flatten the band, generating a HOvHS \cite{NatComYuan}. A plethora of correlated phases are found in this system at partial fillings \cite{lu2019superconductors, andrei2020graphene}, yet, the twisted angle is not an equilibrium state of the bilayer resulting in twist angle disorder across the device \cite{uri2020mapping}. Bernal stacking of bilayer graphene, which is an equilibrium configuration, also hosts a saddle point whose dispersion can be further flattened by application of a displacement field which gaps the adjacent Dirac nodes \cite{zhou2022isospin, de2022cascade, seiler2022quantum}, and similarly in the metastable rhombohedral trilayer graphene \cite{zhou2021superconductivity, guerrero2022rhombohedral}. All these scenarios require either non-equilibrium stacking or the application of external perturbations to generate a HOvHS and thereby induce an electronic instability.

Kagome lattices intrinsically host spectroscopic features with high density of states, as their band structure comprises both a flat band and two sets of saddle points pinned to the $M$ high symmetry momenta at energies above and below a Dirac node pinned at $K$. There are several material realizations of layered kagome metals that display a variety of instabilities or correlated states \cite{yin2022topological}. When the flat band is at the vicinity of the Fermi level heavy-fermion-like phenomenology appears to arise in Ni$_3$In \cite{ye2024hopping} and Mn$_3$Sn \cite{PhysRevLett.125.046401}, and Mott-like phenomenology as in CoSn \cite{chen2024cascade}. Yet, the flatness of this band is not protected, and it can become dispersive under various perturbations, such as the presence of longer-range hoppings. In contrast, the saddle points are more robust and support a variety of spontaneously symmetry broken states, most commonly charge density waves possibly with exotic character, as in the AV$_3$Sb$_5$ family (A=K, Rb, Cs) \cite{neupert2022charge,NatComLuo, NatComHu, NatPhysKang, Scammell2023, nie2022charge} and in FeGe \cite{teng2022discovery,yin2022discovery}. Signatures of nematic state unrelated to charge density wave order have been reported in CsTi$_3$Bi$_5$ \cite{li2023electronic}, and ScV$_6$Sn$_6$ \cite{jiang2024van}. However, their origin in an intrinsic motif of the kagome band structure has remained unclear.

The rise of kagome metals as a host for largely unprecedented avenues of electronic order \cite{neupert2022charge} has brought about a refined understanding of the nature of vHS. Both sets of saddle points in the kagome bands induce logarithmically divergent vHSs in the density of states, however they exhibit distinct orbital character. At the pure (p) type saddle point, the three inequivalent vHS at $M$ exhibit mutually orthogonal sublattice support. The mixed (m) type saddle point exhibits a pairwise inequivalent mixture of two sublattices at each $M$ point. As a consequence, the effect of onsite electronic repulsion within the vHS scattering channels is reduced, termed sublattice interference~\cite{PhysRevB.86.121105}. It has far-reaching consequences for the types of electronic instabilities favorable in kagome materials. For instance, it can naturally explain, from a purely electronic perspective, the observed dominance of charge density wave order over magnetic ordering in kagome metals ~\cite{PhysRevLett.110.126405,PhysRevLett.127.177001}. Yet, most of known kagome metals host $p$-type saddle points at the vicinity of the Fermi level, while material realization of partially filled $m$-type saddle points, and even more so flattened ones featuring a HOvHS, are presently lacking.
Here, we identify surface reconstruction in a kagome metal as a novel mechanism to generate HOvHs in materials under equilibrium conditions compared with existing approaches that are based on parametric tuning~\cite{NatComYuan,PhysRevB.100.121407}. The mechanism is specific to $m$-type saddle points.
We find through scanning tunneling microscopy (STM) and spectroscopic mappings that the direct consequence of this surface HOvHS is the formation of an unconventional nematic charge order that spontaneously breaks rotational symmetry due to a Pomeranchuk instability in the absence of translational symmetry breaking in the form of a charge density wave. By this we show the direct relation between the HOvHS nature of saddle points and the type of electronic instabilities they stabilize opening the path for future exotic symmetry broken states in kagome metals. 

\section{Results}\label{sec2}

\begin{figure}[h]
\centering
\includegraphics[width=0.95\textwidth]{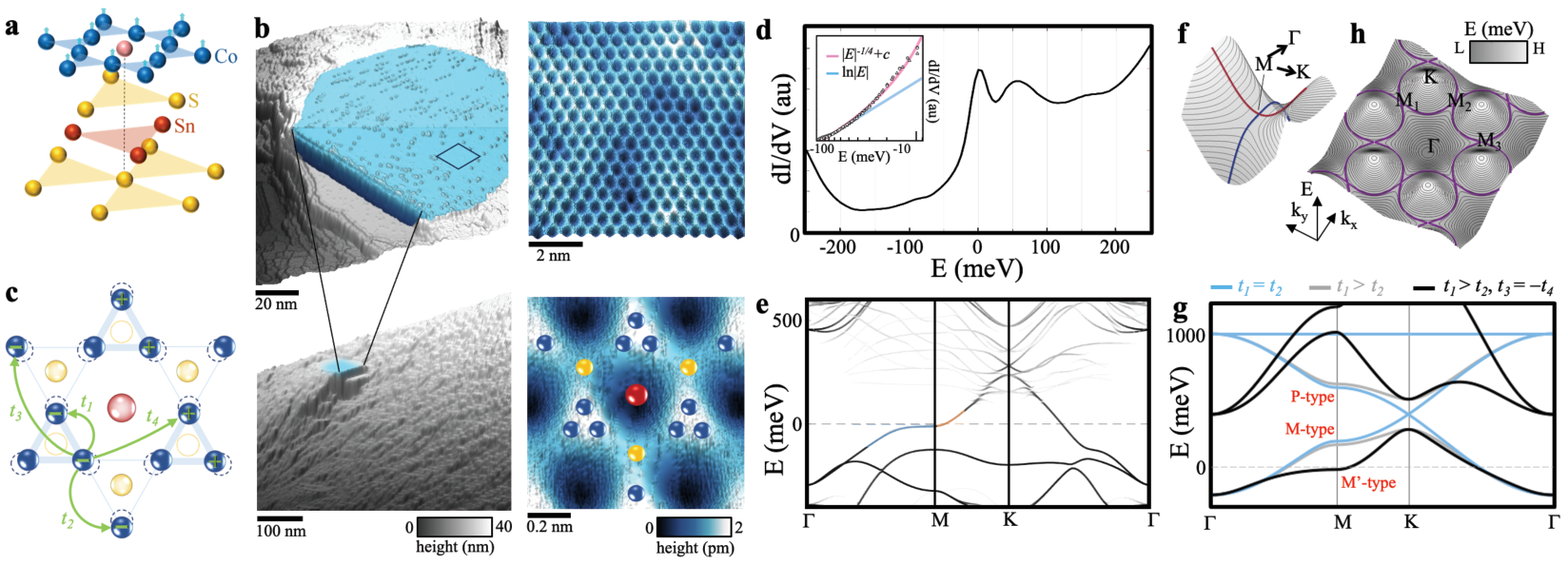}
\caption{\textbf{Higher-order van Hove singularity in a distorted kagome lattice. a} The crystal structure of Co$_3$Sn$_2$S$_2$ with kagome Co arrangement. \textbf{b} Topographic images of the Co$_3$Sn terrace. The highest resolution shows triangular reconstruction. \textbf{c} The reconstruction away from the underlying S ions that explicitly break the symmetry. \textbf{d} dI/dV spectrum showing a zero bias peak. The inset zooms in to the negative bias side of the zero bias peak (circles correspond to the same data set, squares and triangles are taken from two other data sets) with best fits to a logarithm, $\ln{|E|}$ (blue linear line in semi-log plot), and to a power law with constant offset yielding a power of $1/4$ (pink line). \textbf{e} DFT calculation of  Co$_3$Sn$_2$S$_2$ band structure projected to the Co$_3$Sn surface, relaxed by triangular reconstruction. The blue and red segments correspond to the electron and hole-like dispersion around M, respectively \textbf{f} Flat saddle point extracted from DFT and fitted with quadratic and quartic polynomials (red and blue lines, respectively).  \textbf{g} Tight-binding band structure of an unperturbed kagome lattice, kagome with lowest order triangular distortion ($t_1>t_2$), and kagome also with opposite sign inter-unit cell hopping ($t_3=-t_4$) in black, gray and blue lines, respectively.  \textbf{h} Valence band of a distorted kagome lattice with HOvHS at the equi-energy contour denoted by purple line.  }\label{fig1}
\end{figure}

Co$_3$Sn$_2$S$_2$ is a member of the shandite material family and develops ferromagnetic order below 175 K. It is characterized as a time-reversal symmetry broken Weyl semimetal, with Weyl nodes residing about 60 meV above the Fermi energy \cite{morali2019fermi,liu2019magnetic}. The layered crystal structure of Co$_3$Sn$_2$S$_2$ consists of Co$_3$Sn layers in which the Co ions are arranged in a kagome structure (Fig.\ref{fig1}a), and are covalently bonded to triangularly coordinated S and Sn layers. STM measurements require freshly cleaved surfaces, where the common cleave planes expose either the Sn or the S termination. The kagome nature has been previously probed indirectly through the Sn or S termination layers \cite{yin2019negative}. Very rarely, however, a Co termination is exposed, allowing direct investigation of the electronic spectroscopy in presence of the kagome structure. The topographic images depicted in Fig.\ref{fig1}b demonstrate such a termination where a flat Co$_3$Sn terrace appears among a dense sequence of rough step edges. Higher resolution imaging, as shown in the upper right panel, reveals a hexagonal pattern characteristic to kagome surfaces as the individual atomic sites within the corner sharing triangles are typically not revealed. Finer resolution topography, as shown in the bottom right panel, further finds that the $C_6$ symmetry of the kagome Co$_3$Sn layer is broken on the surface down to $C_3$ as three out of the six corner-sharing triangular sites around the hexagon appear slightly elevated relative to the other three. Such a surface reconstruction is rather natural due to the explicit inversion symmetry breaking introduced by exposing the Co termination through cleaving. Indeed, ab initio calculations find that the Co$_3$Sn surface strongly reconstructs by an inward shift of the Co ions by about 10\% of the initial Co-Co bond length within those triangles that do not reside over S ions a layer below, as displayed in Fig.\ref{fig1}c. 

In differential conductance (dI/dV) spectroscopy we find a signature for the existence of a singular density of states at the Fermi level in the form of a robust zero bias peak shown in Fig.\ref{fig1}d. The sharp rise of the density of states on negative energies is better captured by a power law characteristic of a HOvHS than logarithm that signifies regular vHS (inset, pink, and blue lines, respectively). Once we allow a constant offset dI/dV, that may originate from either surface or bulk contributions, the best fit yields a power of $-1/4$ in agreement with a HOvHS with a quartic first non-vanishing expansion coefficient for the dispersion. Excluding such background the fit finds a stronger divergence characterized by a power of $-1/2$ (see Fig.\ref{fig-sm-fits}) suggesting the zero bias divergence is of at least power $-1/4$. Intriguingly, on the Sn termination of Co$_3$Sn$_2$S$_2$ we find quasi-particle interference (QPI) patterns that seem to correspond to a HOvHS at the vicinity of the Fermi energy (see Fig.\ref{fig-sm-snqpi}).

In the calculated ab initio surface-projected band structure, shown in Fig.\ref{fig1}e, we find a single isolated surface band which crosses the Fermi level, and indeed identify a saddle point close to the Fermi level at the $M$ point (details in Sec.\ref{sm-dft}). The dispersion turns from hole-like along the $M-\Gamma$ direction to electron-like along $M-K$ (blue and red segments, respectively), reminiscent of the saddle point of a kagome layer. Since it has no trace in the bulk band structure (see Fig.\ref{fig-sm-dft}a) we conclude it results from a two-dimensional surface state bound at the Co$_3$Sn termination which exhibits the kagome potential. We notice that the hole side of the dispersion along $M-\Gamma$ is very flat, as its energy decreases only by about 20 meV across a third of the Brillouin zone. We fit the dispersion along the $M-\Gamma$ and $M-K$ high symmetry directions with a polynomial momentum dependence, shown in Fig.\ref{fig1}f by blue and red lines, respectively (see also Fig.\ref{fig-sm-polyfit}). While the electron-like dispersion is fairly quadratic, the hole-like dispersion along $M-\Gamma$ has a null quadratic coefficient, a finite quartic coefficient, and demands a dominant sixth power coefficient to accurately capture the flattened dispersion. This formally endows the saddle point as a HOvHS, where the Jacobian and Hessian of the dispersion vanishes, and the Taylor series requires higher powers of momentum beyond quadratic order.

We trace the generation of the HOvHS to the surface reconstruction. We compare the ab initio slab calculations of the surface-projected bands with and without the Co reconstruction, illustrated in Fig.\ref{fig-sm-dft}b. In both settings, we identify a saddle point at the $M$-point. The most prominent change induced by the Co reconstruction is a downward shift in energy of the saddle point level by about 100 meV. Meanwhile, since the bands at the $\Gamma$ and $K$ points are hardly affected, this downward shift results in a pronounced flattening of the hole-like dispersion along $\Gamma-M$ that generates the HOvHS, without any apparent fine-tuning. The Fermi surface develops a pointed, snow-flake shape, near which the density of states diverges as a power law (Fig.\ref{fig:bandstructure}d and e, respectively), both characteristic of a HOvHS. As the saddle point shifts down in energy, the chemical potential becomes pinned to it due to the large density of states. In this way, the surface reconstruction lowers the electronic energy akin to a Peierls instability, where the HOvHS generation plays the role of an effective gapping mechanism by the lattice distortion.

This therefore marks a novel mechanism for generating a HOvHS in kagome metals. We demonstrate its generality by reproducing it in a minimal tight-binding model of a distorted kagome lattice. We position the $d_{z^2}$ orbitals of the Co atoms, which contribute the most to the orbital weight at the saddle point (see Fig.\ref{fig-sm-orbit}), on a distorted kagome lattice. We consider hopping terms up to the fourth nearest neighbor demonstrated by the green arrows in Fig.\ref{fig1}c. Since there are three atoms in the kagome unit cell, this merely amounts to those hoppings which connect adjacent units. The Hamiltonian then reads:
\begin{equation}
\label{tb_model_main}
\mathcal{H} = \epsilon \sum_{i}{c_i^\dagger c_i} -t_1 \sum_{\langle ij \rangle}{c_i^\dagger c_j} - t_2 \sum_{\langle \langle ij \rangle \rangle}{c_i^\dagger c_j} - t_3 \sum_{ \langle \langle \langle ij \rangle \rangle \rangle}{c_i^\dagger c_j} - t_4 \sum_{\langle \langle \langle \langle ij \rangle \rangle \rangle \rangle}{c_i^\dagger c_j}.
\end{equation}
Setting $t_1=t_2$ and $t_3=t_4=0$ reproduces the undistorted kagome band structure displayed in Fig.\ref{fig1}g (blue lines). We next consider the implications of the triangular distortion. At first, we only break the symmetry between nearest neighbor hopping on the distorted triangles by setting $t_1>t_2$ while excluding higher order hopping terms, $t_3=t_4=0$. The main outcome of such distortion, shown by gray lines in Fig.\ref{fig1}g, is the gapping of the Dirac node at the $K$ point due to broken $C_2$ symmetry. We next include symmetry breaking in the higher order hopping terms by introducing a relative $\pi$ phase among them, $t_3=-t_4$. Consequently, the $m$-type saddle point shifts downwards in energy while the bands at the $\Gamma$- and $K$-points are unaffected. We confirm that this results in a considerable flattening of the hole dispersion of the saddle point generating a HOvHS. Fitting the functional dependence of the density of states at the vicinity of the vHS yields algebraic energy dependence with a power of -0.25 (Fig.\ref{fig:bandstructure}b, inset) signifying the higher order nature of the saddle point.

The distorted tight binding dispersion, including the flat band which turns dispersive, all capture well the surface band structure from the ab-initio calculations, overlaid in Fig.\ref{fig:bandstructure}a. We find the analytic dependence of the saddle point dispersion in the tight-binding model, $\varepsilon(\mathbf{k})=a_xk_x^2-a_yk_y^2+b_xk_x^4-b_yk_y^4+b_{xy}k_x^2k_y^2+...$, on the hopping coefficients, $t_1...,t_4$ (see Sec.\ref{sm-model}). Consistent with the polynomial fitting of the DFT, the set of coefficients obtained via Wannierization of the DFT surface bands are very close to those which make the prefactor of the quadratic hole term vanish, $a_y(t_1...,t_4)=0$. We note, that $t_{3,4}$ do not have to be equal in magnitude for the quadratic coefficient of the DOS expansion to vanish (see Fig.\ref{fig:bandstructure}d), but $t_3=-t_4$ seems to fit well the DFT dispersion, and since the hopping distances for both $t_3$ and $t_4$ are similar on the distorted lattice, we expect their magnitudes to be similar. The generation of a HOvHS under such a perturbation is unique to the $m$-type saddle point, and does not develop for the $p$-type one. Consistently, the $m$-type nature may enforce the relative $\pi$-phase of the higher order hopping terms. The pairwise mixing of the unit cell sites corresponds to alternating rows of contributing and non-contributing sites per $M$ point, whereas the $M$ point position at the edge of the Brillouin zone implies a relative sign flip among those rows of contributing sites, marked in Fig.\ref{fig1}c. We thus find a single hopping term will have an opposite sign relative to the other three. Intriguingly, though, when we further examine the site contribution to the HOvHS of the distorted kagome in both the tight binding and the ab initio models, we find that the pairwise mixing turns into an almost equal support from all three unit cell sites (Fig.\ref{fig-sm-polyfit}b). The type of site mixing at the saddle point affects the nature of electronic instabilities which can develop \cite{PhysRevB.86.121105}. The geometry of the Fermi surface is also expected to play an essential role. HOvHS which suppress nesting via the curved Fermi surface edges, visible in the curved electronic pockets (Fig.\ref{fig1}h), accompany a suppressed tendency towards charge density wave instabilities. In such a situation, translationally-invariant symmetry breaking states, such as Pomeranchuk order, are natural instabilities.

\begin{figure}[h]
\centering
\includegraphics[width=0.95\textwidth]{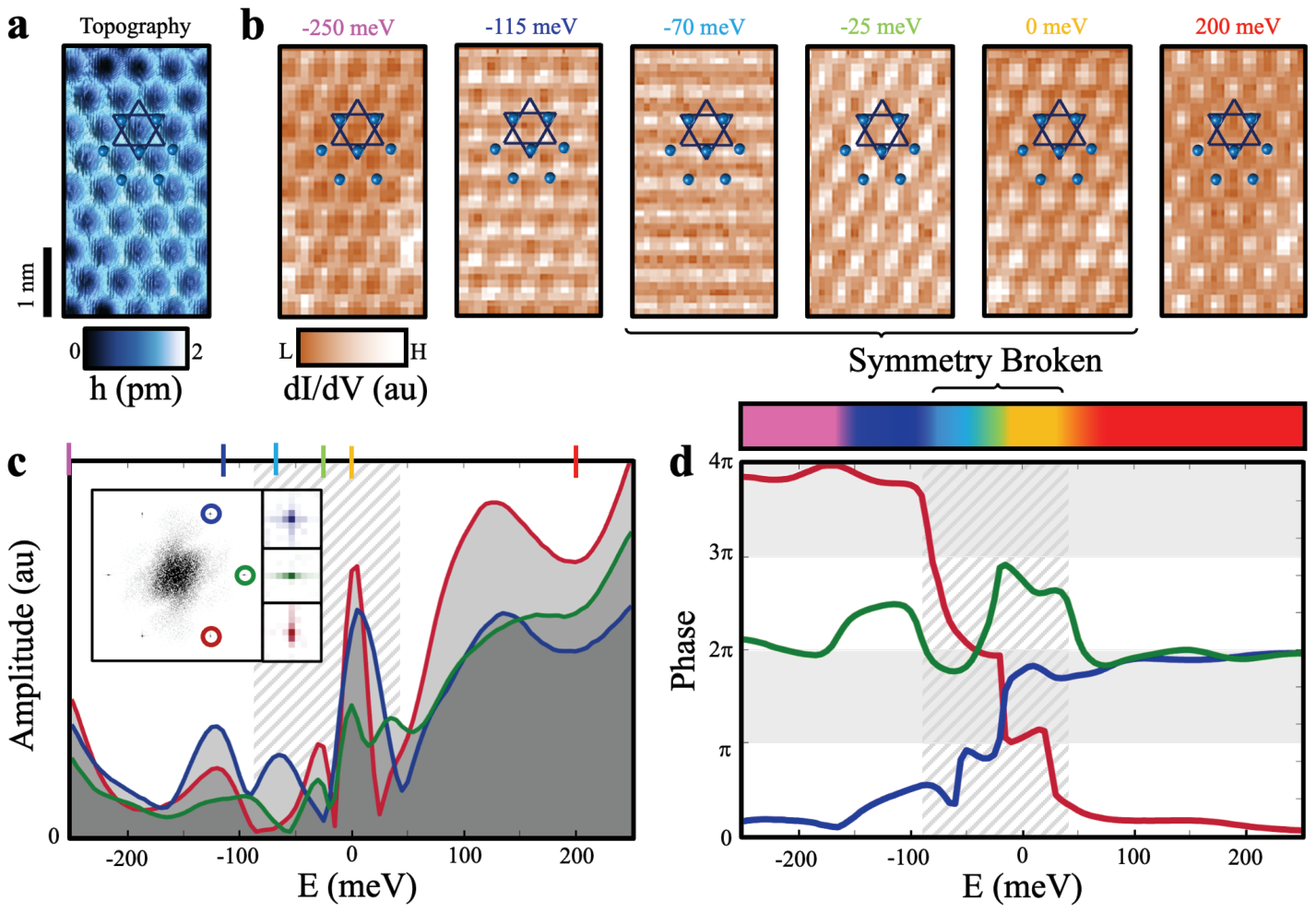}
\caption{\textbf{Cascade of nematic phase shifts. a} Topographic image of the Co$_3$Sn surface with a commensurate kagome lattice overlaid. \textbf{b} dI/dV maps of the same region in a at various energies with the kagome lattice overlaid at a fixed position. \textbf{c} The amplitudes of the three Bragg peaks of the FT dI/dV maps (see one shown explicitly in inset with zoomed panels on the peaks). \textbf{d} Phase evolution of the same three Bragg peaks in energy.}\label{fig2}
\end{figure}

Next, we show that in the presence of the HOvHS at the Fermi energy the surface electrons on the distorted Co$_3$Sn surface exhibit a spontaneous symmetry breaking. We image spectroscopically the spatial distribution of the local density of state across the Co$_3$Sn termination. We pick a region free from any topographically visible surface impurities, as shown in Fig.\ref{fig2}a. We then image the spatial modulation of the differential conductance over that surface with sub-atomic resolution across the bias interval of -250 meV to 250 meV, demonstrated in the panels of Fig.\ref{fig2}b (full field of view of this and five other dI/dV maps taken under various conditions are given in Fig.\ref{fig-sm-fullfov}a-f). At a first glance all these patterns appear similar as they have the same periodicity. Accordingly, no new Bragg peaks are detected in Fourier space. However, when we attempt to position the kagome-coordinated Co atoms we fail to do so in a way that will respect the symmetry of the periodic patterns across all energies imaged. We succeed in doing so consistently only away from the Fermi energy at high and low bias - below about -100 meV and above 50 meV. At all intermediate energies about the Fermi energy we find atomic scale dI/dV distributions that seem to peak at locations that are not symmetry points of the underlying ionic kagome crystal, which remains fixed throughout the measurement.  

Yet, the bare spatial resolution within a single unit cell that the spectroscopic raw mappings provide is marginal for firmly concluding that the electronic structure breaks the symmetry imposed by the ionic lattice. Nevertheless, the extended two-dimensional spatial mapping of hundreds of identical unit cells provides redundancy which we use to demonstrate the presence of nematicity beyond doubt. One method to exploit this is by a Fourier transformation (FT) which resolves the periodic patterns across the extended field of view. We clearly identify the three non-equivalent Bragg peaks corresponding to the unit cell period in the spatial spectroscopic mappings, circled in the inset of Fig.\ref{fig2}c. The right panels show enlarged images of those Bragg peaks in corresponding colors. They consist of a single strong pixel surrounded by a faint cross-like halo that originates mainly from the finite window size, while drift seems negligible. The main panel shows the energy dependence of the amplitudes of the central pixels of the three Bragg peaks in corresponding colors. We find a series of approximate sharp nodes around zero bias at which the amplitude of some or all Bragg peaks almost vanishes. Between those nodes the amplitudes of the Bragg peaks are strongly unbalanced, while beyond this interval they are all finite and fairly equal. We note, that unlike the reproducibility of the nodes and phase shifts at intermediate energies, the slight anisotropy in the magnitude of the Bragg peaks at low and high energies seems to somewhat vary between different spectroscopic maps (see Fig.\ref{fig-sm-brgap}a) and thus do not seem to signify nematicity.

Even more striking though is what we find in the phase information of the FT Bragg peaks, displayed in Fig.\ref{fig2}d. Since the density of states we image is a real positive quantity, its FT is a complex number, $A_\mathbf{q}e^{i\Phi\mathbf{q}}$. Its phase, $\Phi_\mathbf{q}$, entails the phase-shift, $\mathbf{\delta_q}$, with which the periodic pattern of corresponding wavelength $\mathbf{q}$ is positioned in real space, such that $\Phi_\mathbf{q}=q \delta_q$. For the Bragg peaks that we inspect here the $q$ values are the three reciprocal wave-vectors of the kagome crystal. Remarkably, we find at the vicinity of zero bias a sequence of three abrupt phase jumps by a fraction of 2$\pi$ and thus correspond to real space shifts of the electronic wavefunction by a fraction of the unit cell vectors (see a consistent data set in Fig.\ref{fig-sm-brgap}b). As the real space direct mapping hints, at least some of the wave function distributions about the Fermi energy violates the symmetry imposed by the ionic lattice and is thus nematic. Between high and low energies, on the other hand, the phases of the Bragg peaks differ by approximately integer multiples of 2$\pi$, confirming that away from zero bias the wavefunction structure adheres to the underlying ionic lattice. Intriguingly, though, the phases at highest bias do not recover to their values at low bias, as will be discussed later. Every abrupt phase jump is accompanied by a node in the Bragg peak amplitude. We use the phase jumps and nodes to identify five energy intervals, indicated by the colorbar in Fig.\ref{fig2}d and in later figures. 

Yet, in order to directly obtain the exact wavefunction patterns and changes among them with respect to the quenched ionic lattice we aim to achieve superior resolution directly in real space. For this we use the redundancy in mapping multiple equivalent unit cells to construct a super-resolved single unit cell \cite{zeljkovic2012scanning}. It relies on the incommensurability between the ionic lattice, which in the present case has a kagome structure, and the rectangular grid at which the STM measurement probes, as well as the inevitable mismatch in the exact lattice constants of the two. As a result, the unit cell is repeatedly probed at thousands of different points over the periodic unit cell structure which can be collapsed into a single, super-resolved unit cell (more details are presented in Fig.\ref{fig-sm-lattice}).

\begin{figure}[h]
\centering
\includegraphics[width=0.95\textwidth]{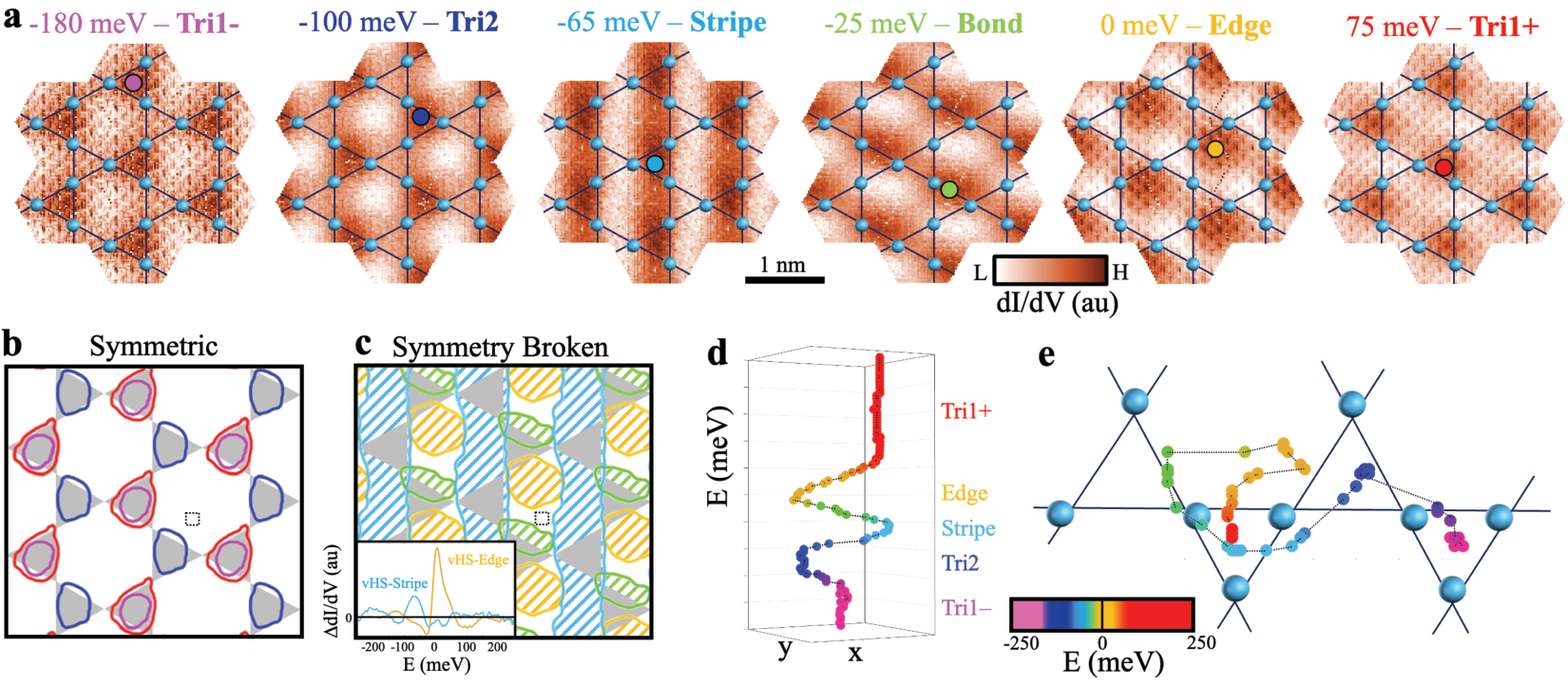}
\caption{\textbf{Super-resolution of spontaneous symmetry broken states. a} Super-resolved cells at the different energies with the kagome lattice overlaid on them. The circles mark the wavefunction's center of mass position. \textbf{b(c)} Contour maps of the wavefunctions from the super-resolved cells at different energies away from (at the vicinity of) the Fermi energy in corresponding color. The inset shows the excess dI/dV spectrum localized within the positions of the nematic inner edge and stripe states (yellow and blue lines, respectively) relative to the extended background (averaged within the dotted square). \textbf{d} Energy evolution of the wavefunction's center of mass position. \textbf{e} Position path of the center of mass across all energies imaged showing a unit cell displacement is acquired.}\label{fig3}
\end{figure}

The results of applying this procedure are presented in Fig.\ref{fig3}a (see the full energy evolution in Fig.\ref{fig-sm-folded} and additional data set in Fig.\ref{fig-sm-fullfov}g). It shows the wavefunction distribution across the super-resolved unit cell at different energies of interest, each tiled six more times to make the spatial periodicity apparent. We indeed confirm with great clarity that the wavefunction distribution at low and high energies, below -150 meV and above 50 meV (left and right panels, respectively), are the same. Above -150 meV and up to about -80 meV the wavefunction shifts within the unit cell to a different position. To compare wavefunction distributions across different energy intervals we extract equi-intensity contours at a value which is 10\% below the maximal one. Such wavefunction contours, representative of the three energy intervals discussed thus far, are overlaid in Fig.\ref{fig3}b. The similar distribution at the highest and lowest energy intervals (red and magenta, respectively) is evident. The contours from the adjacent negative energy interval (blue contours) reside over sub-surface S sites, complementing them to a kagome pattern. We thus use these unperturbed wavefunction distributions to position the underlying kagome lattice (balls and sticks in Fig.\ref{fig3}a and gray triangles in Fig.\ref{fig3}b and c) in agreement with the symmetry of the topographic image. 

Remarkably, the remaining energy interval closest to the Fermi energy of the super-resolved unit cells reveals three distinct wavefunction distributions that do not respect the determined ionic lattice symmetry, in accordance with the FT phase analysis (Fig.\ref{fig2}d). Within the energy interval -80 to -35 meV we find a stripe-like wavefunction distribution (see also Fig.\ref{fig-sm-fullfov}). Picking one stripe orientation over the other two constitutes a spontaneous symmetry breaking of the $C_3$ rotation. We note, that indeed within this energy interval the amplitude of a single Bragg peak dominates while the other two almost vanish as shown in Fig.\ref{fig2}c. About the Fermi energy and within the interval of -20 to 50 meV the wavefunction localizes at the inner edge of the central hexagon of the kagome pattern. Again, this location is equivalent to the two other edges related by $C_3$ symmetry and thus hallmarks spontaneous breaking of the $C_3$ symmetry of the original kagome lattice. In the narrow energy interval between -35 to -20 meV, the super-resolved unit cell shows a distinct distribution which seems to be localized over a Co-Co bond and again breaks the $C_3$ symmetry. The narrow energy interval over which we observe the bond distribution may suggest it is an intermediate configuration between the edge and stripe states at higher and lower energy intervals. 

The equi-density contour lines of the symmetry broken nematic states within the distorted kagome unit cell are displayed in Fig.\ref{fig3}c. The location of the underlying kagome ions, fixed at high and low energies, are given by the gray triangles. The symmetry broken states seem to occupy different inner edges of the hexagonal region at different energy intervals. It also becomes clear that no position of the kagome ionic lattice will be commensurate with the wavefunction distributions across all energies. We rule out strain as the mechanism behind the symmetry broken wavefunction distributions because a distorted lattice results in distorted electronic states at all energies (as we demonstrate via ab initio calculations, see Fig.\ref{fig-sm-strain}) while the symmetry breaking we observe is limited to the vicinity of the Fermi energy. We also rule out tip effects in inducing the nematic distributions, by mapping the spectroscopy across the Co$_3$Sn to S step-edge in Fig.\ref{fig-sm-svsco}, showing that the latter does not exhibit nematicity while the former does.

The super-resolved cells enable us to closely follow the exact redistribution of the wave function with energy beyond the representative snapshots displayed in Fig.\ref{fig3}a. For this, we calculate the `center of mass' of the wavefunction as a function of energy:
$\mathbf{r}_{com}=\int_{uc} d\mathbf{r'} |\Psi(E,\mathbf{r'})|^2 \mathbf{r'}$. For example, the center of mass locations calculated for the snapshots in Fig.\ref{fig3}a are given by the solid circles, while their full energy evolution is shown in Fig.\ref{fig3}d. The distribution of the unoccupied states as well as that of core energies are completely quenched showing no energy dependence. Only a thin shell of about 100 meV around the Fermi level of mainly occupied states shows rapid redistribution. The edge and stripe states (yellow and light blue, respectively) are stable over finite energy intervals. In contrast, the intermediate bond state (green) indeed seems more as a transient between them. 

Following accurately the wavefunction `center of mass' progression across the distorted kagome unit cell, as shown in Fig.\ref{fig3}e, reveals another remarkable property that is not apparent through the individual snapshots. We find that the high- and low-energy distributions, which at first glance appear identical, rather accumulate a relative shift of a full unit cell vector. This unit cell shift is in accord with the relative 2$\pi$ phase offsets of the Bragg peaks between high and low energies we have observed in Fig.\ref{fig2}d. It signifies an evolution of the wavefunction with energy akin to `charge pumping'. We stress, though, that in STM we do not gate the sample, so actual charge pumping in its colloquial sense does not take place here. Still, we speculate it may have an effect on boundaries at step edge terminations, where we have indeed previously reported the appearance of an edge accumulation at about -100 meV \cite{howard2021evidence}. 

\begin{figure}[h]
\centering
\includegraphics[width=0.95\textwidth]{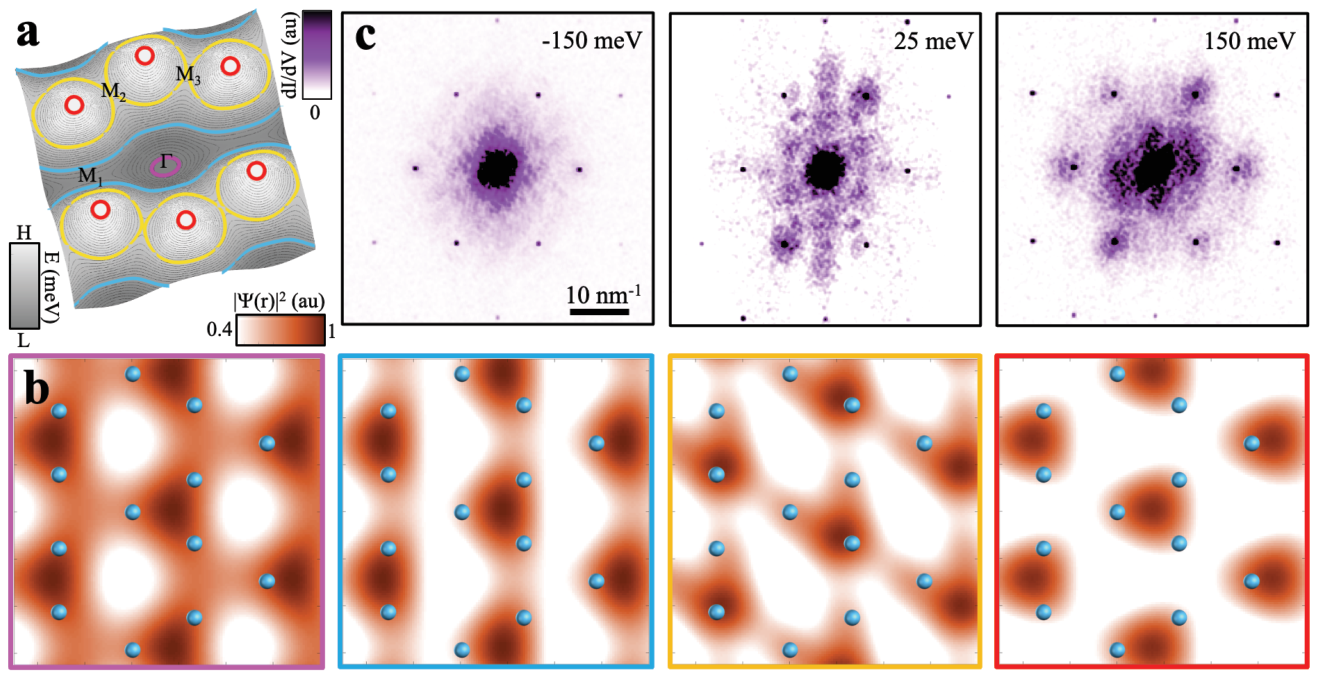}
\caption{\textbf{d-wave Pomeranchuk instability to a nematic state. a} Calculated kagome bandstructure of the tight-binding Hamiltonian Eq. \eqref{tb_model_main} perturbed by the d-wave Pomeranchuk order parameter. \textbf{b} Local density of states calculated using the perturbed tight-binding Hamiltonian, ranging from high to low energies (frame color corresponds to equi-energy contours in a). The lattice sites are denoted by blue spheres. At low and high energies, the density of states respects the lattice symmetry, while close to the Fermi energy we find nematic distributions (middle panels). \textbf{c} QPI showing $C_3$ symmetry broken pattern at the vicinity of the Fermi energy (middle  panel) and symmetric away from it (left and right panels)}\label{fig4}
\end{figure}

We thus identify two nematic electronic states, one at the Fermi energy and one about 60 meV below it, that spontaneously break rotational symmetry of the underlying lattice but do not break its translational symmetry. We attribute the appearance of the two nematic states to a $d$-wave Pomeranchuk instability. With the onset of Pomeranchuk order, the distorted kagome bandstructure further deforms such that saddle points at a subset of the $M$ points are pushed down away from the Fermi energy, demonstrated in Fig.\ref{fig4}a by comparison of $M_1$ with $M_2$ and $M_3$, spontaneously breaking the threefold rotational symmetry of the distorted kagome lattice. To describe the interacting instabilities of this theory, we develop a patch model that incorporates the degrees of freedom located near the three $M$-points where the density of states is largest (see Sec.\ref{sm-model}). The interacting model of a HOvHS is equivalent to one developed previously \cite{Classen2020}, though the renormalization group equations themselves differ slightly due to the fact that Co$_3$Sn$_2$S$_2$ is a ferromagnet, and so there is no spin degeneracy in the present case. With the assumption that the largest interaction matrix elements are between states with small momentum transfer, $d$-wave Pomeranchuk order appears already at mean-field level. We find moreover that $d$-wave Pomeranchuk order remains the leading instability of the model upon including corrections beyond mean-field, introduced by integrating out the degrees of freedom away from the Fermi surface producing a renormalization group flow. 

The resulting ground state is a linear combination of two $d$-wave form factors $\mathcal{O}_{1}(k_x^2- k_y^2) + \mathcal{O}_2 2k_xk_y$. The fact these states are translationally symmetric charge orders means the Landau free energy possesses a cubic term $\mathcal{F}_3=O_1^2-3\mathcal{O}_1 \mathcal{O}_2^2$, resulting in a minimum of the free energy in which $O_{1,2}$ are nonzero at all three $M$-points, but largest at one particular spontaneously chosen $M$-point, producing a charge nematic state as shown in Fig.\ref{fig4}a. 
Adding such a Pomeranchuk order parameter to the tight-binding model of Eq. \eqref{tb_model_main}, we calculate the resulting local density of states at different energies (corresponding to equi-energy contours in a) shown in the panels of Fig.\ref{fig4}b. At intermediate energies, we indeed recover a nematic wavefunction distribution following from the $d$-wave Pomeranchuk distortion of the Fermi surface. The one at lower energies resembles a stripe distribution and the one at higher energies localizes at a point that breaks the rotational symmetry of the lattice. At high and low energies, the wavefunction distribution becomes symmetric as we find in experiment. While our accompanying theoretical investigation suggests a $d$-wave Pomeranchuk instability, further theory refinement will be necessary to rationalize all detailed features of the nematic phase we observe.

Finally, we carefully examine QPI patterns acquired over regions with visible adatom impurities that act as scatterers. As the scattered electrons interfere they embed the wavelength of the transferred momentum, $\mathbf{q}$, in the LDOS that can be resolved by FT. Such FT images are presented in Fig.\ref{fig4}c below, close to, and above the Fermi energy. Remarkably, close to the Fermi energy we identify among the fairly symmetric QPI patterns an additional pattern dispersing radially along a single $\Gamma-M$ direction.  We attribute the absence of the $C_3$ symmetric counterparts to the removal of saddle points from the Fermi energy that eliminates those elastic scattering processes. The QPI pattern goes symmetric away from the Fermi energy as demonstrated in Fig.\ref{fig4}c (the full energy evolution and an additional data set are given in Fig.\ref{fig-sm-coqpi}).  

In summary, we have identified a mechanism by which distorted kagome lattices develop HOvHS at their $m$-type saddle point, realized on the Co$_3$Sn surface of ferromagnetic Co$_3$Sn$_2$S$_2$, where the HOvHS is pinned at the Fermi energy. The manifestation of this effect is that the Fermi surface deforms, undergoing a Pomeranchuk instability by spontaneously displacing a subset of the three distinct saddle points to lower energies. As a result, the electronic states become nematic, and the wavefunctions break the symmetries of the underlying distorted kagome lattice, which we trace both in real space through the construction of super-resolved unit cells, and in reciprocal spaces through the phase and amplitude of the Bragg peaks. Our observations open the path for further explorations of HOvHS generation in other distorted kagome materials, as well as its three dimensional counterpart the pyrochlore lattice, wherein other forms of instabilities may be established.

\newpage

\section{Figures}\label{sec6}

\newpage{}
\clearpage

\newpage
\clearpage

\newpage{}
\clearpage

\newpage{}
\clearpage

\bibliography{sn-bibliography}
\end{document}